\documentstyle[11pt]{article}
\def\v1{\vspace{1cm}}
\def\be{\begin{equation}}
\def\ee{\end{equation}}
\def\bc{\begin{center}}
\def\ec{\end{center}}
\def\ik{\partial}
\def\vh{\varphi}
\def\ve{\varepsilon}

\newcommand{\bt}{\begin{tabular}}
\newcommand{\et}{\end{tabular}}

\newcommand{\bea}{\begin{eqnarray}}
\newcommand{\eea}{\end{eqnarray}}

\begin{document}

\title
{\bf Hamiltonian Reduction of General Relativity and Conformal Unified Theory}

\author{
M. Pawlowski\thanks{Soltan
Institute for Nuclear Studies, Warsaw, Poland.;
e-mail: pawlowsk@fuw.edu.pl},
V. N. Pervushin\thanks{Bogolubov Laboratory on Theoretical Physics,
Joint Institute for Nuclear Research, Dubna, Russia},
V. I. Smirichinski\thanks{Bogolubov Laboratory on Theoretical Physics,
Joint Institute for Nuclear Research, Dubna,
 Russia;
 e-mail: smirvi@thsun1.jinr.ru} }
\maketitle

\date{\empty}

\begin{abstract}

We discuss the application of the method of the gaugeless Hamiltonian
reduction to general relativity. This method is based on explicit resolving
the global part of the energy constraint and on identification of one of the
 metric
components with the evolution parameter of the  equivalent unconstrained
(reduced) system.

The Hamiltonian reduction reveals a possibility to unify General Relativity
and Standard Model of strong and electro-weak interactions
with the modulus of the Higgs field identified with
the product of the determinant of 3D metric and the Planck constant.

We give the geometrical foundation of the scalar field, derive and discuss
experimental consequences of this unified model:  the cosmic Higgs vacuum,
the Hoyle-Narlikar cosmology,
a $\sigma$-model version of Standard Model without
Higgs particle excitations and inflation.

\end{abstract}

\section{Introduction}

An identification of physical quantities in General Relativity (GR) is a
long-time problem which stimulated Dirac to formulate, for this aim, the
general Hamiltonian theory for constrained systems \cite{d} developed
later by many authors (see e.g. monographs \cite{ks,fs,ht}).

Main difficulties of this identification are the mixing of the
parameters of general coordinate transformations with the true dynamic
variables and the problem of gauge ambiguities.  In the last years there
appeared some new ideas to overcome these difficulties.

The first idea is the idea of the gaugeless Hamiltonian reduction.  It
belongs to Shanmugadhasan \cite{s} (see also \cite{gkp,gkm}) who paid 
attention to the fact that, in
the theory of differential equations with partial integrals of the type of
the first class constraints, Levi-Civita \cite{lc} managed to remove
"gauge ambiguities" by explicit resolving "constraints" without any
additional second class constraints of the type of gauge fixing in the Dirac
terminology \cite{d}.  Levi-Civita used canonical transformations
to convert the first class constraints into new superfluous
momenta\footnote{Recall that Faddeev \cite{f} used canonical
transformations to convert additional second class constraints(gauges)
into momenta to prove the Faddeev-Popov functional integrals \cite{fp}.}.
In this case, the Hamiltonian of the system on constraints automatically
does not depend on the corresponding superfluous variables and gauge
fixing is not needed.

The second idea is the idea of classification of times of the Hamiltonian
reduction \cite{k}.  It follows from the application of the Levi-Civita
prescription to the system with invariance under the time
reparametrization transformations
\cite{kpp,kppp,grg,plb}.
The explicit
resolution of the energy-type constraints shows that one of the former
(superfluous) variables of a constrained (extended) system leaves the
extended phase space to become the evolution parameter of the reduced
system.  This is just the variable with negative contributions to the
energy-type constraints.  In addition, every action of relativistic theory
has to be supplemented by a geometrical convention which relates a
measurable invariant interval with parameters and variables of the
extended system.  One of the variables plays the role of the Lagrange
multiplier in the Dirac general Hamiltonian description.  Thus, we face
with three "times" of the Hamiltonian reduction of the constrained
extended system.  The first is the coordinate time in the initial extended
action.  The second is the evolution parameter of the reduced system
which is one of former variables of the extended system.  The third time
is a product of the first coordinate time and the Lagrange factor (or the
lapse function in the Dirac-ADM foliation of the metric in GR).
In Special Relativity (SR) the
third time it is the proper time of an observer.
 The gaugeless Hamiltonian reduction represents explicit resolving of the
energy-like constraint.  As a result, the extended system is reduced to the
subsystem where the role of evolution parameter is played by the
superfluous variable.  In addition, we have "proper time dynamics"
described by two equations of ES for the superfluous variable and its
momentum.  In SR, the proper time dynamics is nothing
but the relationship between the proper time of an observer and the proper
time of a particle (i.e. the Lorentz transformation).  In cosmological models,
the proper time dynamics is just the Friedmann-Hubble law of evolution of the
Universe.  Here the role of superfluous variable with a negative
contribution to the energy constraint is played by the cosmic scale factor.
The Hubble law is a consequence of the dependence of the proper time of
a comoving observer on the evolution parameter, the cosmic scale factor.

In this classification of times, the main problem of the
Hamiltonian reduction in GR is to pick out the global variable which can
play the role of the evolution parameter of the corresponding reduced
system.
The third idea is the idea of identification of this evolution parameter
with the  global component of the determinant of
the 3D metric of the Dirac-ADM foliation of the space time \cite{grg,plb}. We
can extract this parameter by solving the global part of the energy constraint
where the role of the superfluous momentum is played by the second fundamental
form. Note that the idea to consider the trace of the second fundamental form
as the time-like variable was discussed in \cite{yk,mtw}.

To represent 3D space metric determinant as an independent variable of
 the extended
system, one should use the so - called conformal invariant Lichnerowicz
variables \cite{l} that are suited for studying the problem of initial data
in GR \cite{yk}.  The Hamiltonian reduction in terms of Lichnerowicz
variables reveals a possibility to unify
General Relativity and the Standard Model with the modulus of Higgs scalar
field identified with the product of the determinant of 3D metric and the
Planck constant.  The obtained unified theory is described by the
Lagrangian without any dimensional parameter and it is conformally invariant.
In this Conformal Unified Theory (CUT) \cite{pr,grg,plb}, 
 the observer can measure
only conformally invariant observables including the corresponding interval
(\ref{dscut}) \cite{grg}.  Although gravitational parts of GR and CUT
are mathematically equivalent, they differ by the chosen physical
convention about measurable quantities.
The geometrical foundation of this theory can be obtained in the scalar
version of the Weyl geometry of similarity \cite{HW} where a
scalar field is the measure of change of the length of a vector in its
parallel transfer (like, in the Riemann geometry, the metric is the measure
of change of the direction of a vector).

The present paper is devoted to the discussion of the gaugeless Hamiltonian
reduction in General Relativity and to the derivation of
physical consequences of
Conformal Unified Theory.  The content of the paper is the following:
In Section 2,  we recall the method of the Hamiltonian reduction using as
examples classical mechanics, special relativity, and cosmological models
with invariance under the time reparametrizations.  We give the
classification of times of the Hamiltonian reduction.  Then, in Section 3,
we discuss the extraction of the global variable from the metric in GR.
Section 4  is devoted to Conformal Unified Theory.

~
\section{Hamiltonian reduction}

\subsection{Classical mechanics}

The main problem of General Relativity theory is a complete separation
of true physical variables from parameters of general coordinate
transformations.

The simplest example of a general coordinate transformation is
reparametrization of time.  To formulate the problem, we will first
consider the version of classical mechanics with time reparametrization
invariance.

To get this example, we start with a generic classical system given by the
Hamilton action:

\be\label{rsc}
W^{RS}\left[p_i,
q_i|q_0\right]=\int\limits_{q_0(1)}^{q_0(2)}
dq_0\left[\sum\limits_ip_i
\frac{dq_i}{dq_0}-H^{RS}(p_i, q_i)\right].
\ee

The superscript "RS" means a "reduced system"; it was introduced in order to
distinguish it from the "extended system" (ES) defined below.  Here we
have time $q_0$ which is the{ \sl evolution parameter}.  The system is
invariant with respect to the displacement in time $q_0\to q_0+\delta$ (as
$H^{RS}$ is $q_0$ - independent) but it is not invariant with respect to
the general time reparametrization $q_0\to q_0^\prime(q_0)$.

An extended reparametrization - invariant system can be constructed if we
introduce a "superfluous" pair of canonical variables ($p_0,q_0$) and the
Lagrange factor N:

\be\label{esc}
W^{ES}\left[p_i,
q_i; p_0,
q_0|t, N\right]=\int\limits_{t_1}^{t_2}dt\left[\sum\limits_ip_i \dot q_i-
p_0\dot q_0 -NH^{ES} \right]
\ee
 where
$$
H^{ES}(q_0,p_0,q_i,p_i)=[-p_0+H^{RS}(p_i, q_i)]
$$
is the extended Hamiltonian.

The {\sl coordinate time} $t$ has been introduced.  The extended system is
invariant with respect to its general reparametrization $t\to
t^\prime=t^\prime(t)$.

The {\sl Hamiltonian reduction} means explicit solution of the
equations for the "superfluous" variable:

$$
\frac{\delta W^{ES}}{\delta N}=0\,
\Rightarrow\,
-p_0+H^{RS}(p_i, q_i)=0,
$$

$$
\frac{\delta W^{ES}}{\delta q_0}=0\,\Rightarrow\, \frac{dp_0}{dt}=0,
$$

\be\label{despc}
\frac{\delta W^{ES}}{\delta p_0}=0\,
\Rightarrow\, dq_0=Ndt.
\ee

The first of them is a constraint, the second is a conservation law.  If
we substitute the solution of this equation into
the extended action (\ref{esc}), we get
the conventional action for classical mechanics (\ref{rsc}) with the
parameter of evolution as a former variable of ES.

The third equation for the superfluous momentum $p_0$ (\ref{despc}) is the
relation between the evolution parameter $q_0$ and a combination of the
Lagrange factor $N$ and the coordinate time $t$. If we introduce the
notion of {\sl proper time} $T$

\be\label{ptdef}
dT \equiv Ndt,
\ee
then equation (\ref{despc}) converts into the {\sl proper time dynamics}
(PTD)
$$
dT=dq_0.
$$

The proper time dynamics relation (\ref{ptdef}) is very simple
in the present case of a classical - mechanics system.  It is not the case,
in general, as we will see below in SR and cosmology.  There, instead of
"coincidence", we get the Lorentz transformation for SR and the Hubble law
for cosmological models.  Our aim is General Relativity.

The equation for the "superfluous" canonical momentum establishes the relation
between the evolution parameter $q_0$ of RS and invariant time $T$
constructed with the use of the Lagrange factor.

Thus we have three times of the Hamiltonian reduction:

i) the coordinate time $t$,

and the two times that are parametrized by the former one and are
reparametrization - invariant:

ii) the evolution parameter $q_0$ of RS as the former variable of ES and

iii) the proper time $T$ defined by (\ref{ptdef}).

In the general case, in the process of Hamiltonian reduction, any extended
system

$$
W^{ES}\left[p_i,
q_i; p_0,
q_0|t, N\right]=\int\limits_{t_1}^{t_2}dt\left(-p_0\dot q_0+\sum\limits_ip_i
\dot q_i-NH^{ES}(q_0,p_0,q_i,p_i) \right)
$$
is split into two parts. The first part is a set of reduced subsystems

$$
W^{RS}_{(1,2,...)}\left[p_i,
q_i|q_0\right]=
\int\limits_{q_0(1)}^{q_0(2)}dq_0\left[\sum\limits_ip_i
\frac{dq_i}{dq_0}-H^{RS}_{(1,2,...)}\right]
$$
corresponding to the set of different solutions of the energy constraint

$$
H^{ES}=0 \Rightarrow {P_0}_{(1,2,...)} =
{H^{RS}}_{(1,2,...)}.
$$
The second part is given by the "proper time dynamics" determined by
the equation for superfluous momentum

$$
\frac{\delta W^{ES}}{\delta p_0}=0\,
\Rightarrow\, \frac{dq_0}{dT}=-\frac{\partial H^{ES}}{\partial p_0}
\buildrel{\rm def}\over= \sqrt{\rho(q_0)} \Rightarrow\,
T(q_0)=\int\limits_0^{q_0}\frac{dq_0}{\sqrt{\rho(q_0)}}.
$$

This is the evolution of the proper time [given by the relation
(\ref{ptdef}) which is the chosen convention]
with respect to the evolution parameter of the reduced system.

There are two facts of the considered gaugeless Hamiltonian 
reduction worth to emphasize:

I.  The evolution parameter $q_0$ of RS is one of the initial variables of
ES.

II. Variation principle is added by the convention about measurable time.

As will be shown later, we can change the action independently keeping the
convention unchanged or we can change the convention keeping the action
fixed.

\subsection{Proper time dynamics in special relativity}

Here we will present two examples of the proper time dynamics.
The first of them is given by SR. The Hamiltonian action reads

\be\label{essr}
W^{ES}\left[p_i,q_i; p_0,q_0|t, N\right]=
\int\limits_{t_1}^{t_2}dt\left(-p_0\dot q_0+\sum\limits_ip_i
\dot q_i-\frac{N}{2m}[-p^2_0+H_c]\right)
\ee
and the proper time of an observer is given by
$$
dt=Ndt
$$
according to our convention (\ref{ptdef}) and to the ordinary description.

Variation of (\ref{essr}) with respect to $p_0$ leads to the proper time
dynamics:

$$
\frac{\delta W^{ES}}{\delta p_0}=-{dq_0\over dt}+N{p_0\over m}=0,
\Rightarrow\,
T(q_0)_{\pm}=\pm\int\limits_0^{q_0}dq_0\sqrt{\frac{m^2}{p_i^2+m^2}} =
\pm q_0\sqrt{1-v^2}.
$$

If we recall the relation between $p$ and $v$ in SR, we get the relation
between the proper time and the evolution parameter.  This relation
represents the Lorentz transformation of the proper time of a particle into
the proper time of an observer.
\bigskip

\subsection{Proper time dynamics in cosmology}

The time reparametrization - invariant cosmological models follow from the
Einstein-Hilbert action for the FRW metric with a constant 3-D curvature

$$
W=\int d^4x\sqrt{-g}\left[ -\frac{\mu^2}{6}{}^{(4)}R(g) +
{\cal L}_{\rm mat}\right]; ~~~~~ {}^{(3)}R(\gamma^c)=\frac{6k}{r_0^2};
~~~\mu=M_{\rm Pl}\sqrt{3/8\pi}
$$
Here, the role of evolution parameter
is played by the cosmic scale factor $a_0$:
\be\label{dscos}
ds^2 =
a_0^2(t)[N_c^2dt^2-\gamma^c_{ij}dx^idx^j], ~~~~~~~~ds|_{dx=0}=a_0dT_c=dT_F .
\ee
where $T_F$ is the FRW time and $T_c$ is the conformal proper time
introduced according to our convention.

With the above notation the action of  ES is given by

\be\label{escos}
W^{ES}\left[p_f, f; p_0,
a_0|t, N_c\right]=\int\limits_{t_1}^{t_2}dt\left(-p_0\dot
a_0
+\sum\limits_fp_f \dot f-N_c\left[-\frac{p^2_0}{4V_0}+H_c
\right]\right )
\ee
where
$$
H_c(a_0,p_f, f)=
-V_0 K a^2_0+H_{\rm mat}(p_f, f), ~~~~~~K=\mu^2kr_0^{-2}
$$

The proper time dynamics (or the equation for superfluous momentum) describes
the evolution of proper time of an observer (\ref{dscos})
with respect to the evolution parameter, which is the scale factor

\be\label{ptdcos}
\frac{\delta 
W}{\delta p_0}=0\,
\Rightarrow\,
\frac{da_0}{dT_c}=\pm \rho^{1/2}
\Rightarrow
T_c(a_0)_\pm=\pm\int\limits_0^{a_0}da\rho^{-1/2}.
\ee
where $\rho$ is the density of matter in cosmological models
$$
\rho=\frac{H_c}{V_0} =
\frac{\rho_{\rm anisotropic}}{{a_0}^2}+\rho_{\rm rad}+\rho_{\rm dust}{a_0}
-{\rm K}{a_0}^2+{\rm \Lambda}{a_0}^4.
$$

Inverting the PTD relation (\ref{ptdcos}) we get the Hubble law which reflects
the evolution of scale factor:

\be\label{hubble}
a_0=a_0(T_F)\Rightarrow Z=\frac{a_0(T_F-{D/c})}{a_0(T_F)}-1\simeq ({D/c})
 {H_{\rm Hub}}(T_F)
+\dots.
\ee

While a photon emitted by a star atom some time ago is flying, the Universe
is expanding with all lengths, including the wave-length of this photon,
which becomes more red than a photon emitted by the same standard atom on
the Earth.

Afterwards we'll show that the convention about measurable
time can be changed and another cosmological picture with a similar Hubble law
can be obtained.

\subsection{Gaugeless Hamiltonian reduction and quantum cosmology}

To demonstrate the facilities of gaugeless Hamiltonian reduction, let us
consider quantum cosmology using as an example the Universe filled with
photons.

We have the extended action (\ref{escos}) with the matter Hamiltonian
$H_{\rm mat}$ given by
$$
H_{\rm mat}=H_{\rm photon} = \sum\limits_{K}\frac{1}{2}(p_K^2
+\omega_K^2q^2_K)
$$

The PTD based on convention (\ref{dscos})
and the constraint $H^{ES}=0$
describes the Hubble law for the radiation stage.

The same constraint $H^{ES}=0$ can be used to get the WDW equation in the
conventional approach to quantization.

$$
\left(-(\frac{\hat P_0^2}{4V_0}+V_0k\frac{\hat a_0^2}{r^2_0})
+\hat H_{\rm photon}\right)\Psi_{WDW}=0 ~~~ ~~~~
\left(\hat P_0=\frac{1}{i}\frac{d}{da_0}\right)
$$

There arise the questions : what is the physical interpretation of the wave
function $\Psi_{WDW}(a)$, what is the reason of its nonnormalizability, what
is its connection with the Hubble law $T(a)$?

To reply all these questions, it is enough to recall the convention for the
proper time (\ref{dscos}) and to make the Levi-Civita canonical transformation
\cite{lc,s,gkp} which converts the
constraint into a new momentum

$$
(P_0,a_0) \Rightarrow (\Pi,\eta);
\;\;\;\;\frac{P_0^2}{4V_0}+\frac{ka_0^2}{r_0^2}V_0=\Pi.
$$
so that the new scale variable (as the new
parameter of evolution) coincides with the proper time (like in classical
mechanics).

In this case we have two maps of the canonical transformations and two
Universes.

The action in terms of the new variables reads
$$
W^{ES}_{\pm}=\int\limits_{t_1}^{t_2}dt\left[
\sum\limits_{K}p_K\dot q_K\pm\Pi\dot\eta-
N_c
(-\Pi+H_{\rm  photon})\right].
$$
We get the constraint
$$
\frac{\delta W^{ES}_{\pm}}{\delta N_c}=0
\Rightarrow
\Pi = H_{\rm photon}
$$
and the simplest proper time dynamics
$$
\frac{\delta W^{ES}_\pm}{\delta \Pi}=0
\Rightarrow
d \eta = \pm N_c dt=\pm dT
$$

In this version the WDW equation coincides with the Schr\"odinger equation
$$
\frac{d}{id\eta}\Psi_{HR}(\eta|q)=H_{\rm photon}
\Psi_{HR}(\eta| q);\;\;\;\; \hat\Pi=\frac{d}{id\eta}
$$
and we get \cite{kpp,kppp} the spectral
decomposition over normalizable eigenfunctions
\be\label{spectral}
\Psi_{HR}(\eta|q)=\sum\limits_E\left[
e^{iET}<E|q>\alpha^{(+)}+
e^{-iET}<E|q>^*\alpha^{(-)}\right]=\sum\limits_{E}\Psi_E(T);
\ee
$$
H_{\rm photon}<E|q>=E<E|q>;  ~~~~~~ \Psi_{WDW}\neq \Psi_{HR}.
$$
The obtained wave function (\ref{spectral}) can be
 simply interpreted as the wave function of photons in a
box and it bears a direct relation to the Hubble law.
The derivative of each term of the spectral decomposition
with respect to the
measurable time $T_F$
(\ref{dscos}) gives the measurable energy of the red shift (\ref{hubble})
$$
\frac{d}{idT_F}\Psi_E=\frac{E}{a(T_F)}\Psi_E.
$$
The wave function becomes normalizable as one of variables ($\eta$) leaves
the phase space.

\bigskip

\subsection{Statement of  problems}

We will formulate the statement of problems using SR as an example.

In SR, the reduced Hamiltonian is a square root of the sum of squares of
the momentum and mass. The fundamental question of  GR is:
what is the reduced Hamiltonian of this theory?

In SR we have the reduced action with the evolution parameter as a former
dynamic variable of ES.  What is the evolution parameter in GR? Is it the scale
factor, or the second fundamental form or something else? What is the proper
time dynamics of GR?

If we fulfill the Hamiltonian reduction program in SR, we can quantize and
perform the
spectral decomposition of the normalizable wave function. What are similar
quantization, spectral decomposition and normalizable wave function in GR?
\bigskip

\section{General Relativity and Conformal Unified Theory}

We will to present our version of solving these problems in GR.

GR is based on the Einstein-Hilbert action
\be\label{esgr}
W^{ES}[g_{\mu\nu},F]=\int d^4x\sqrt{-g}\left[
-\frac{\mu^2}{6}{}^{(4)}R(g) +
{\cal L}_{\rm mat}(g,F)\right ],
~~~~~~
\mu=M_{\rm Pl}\sqrt{3/8\pi}.
\ee
and the convention about measurable interval
\be\label{dsgr}
(ds)^2=g_{\mu\nu}dx^\mu dx^\nu.
\ee

Both the action and the convention are invariant under general 
coordinate transformations.

For the Hamiltonian description one conventionally
uses the Dirac-ADM foliation of four-dimensional metric with pointing out the
time is of an observer

\be\label{adm}
ds^2=N^2 dt^2- {}^{(3)}g_{ij}\breve{dx}{}^i
\breve{dx}{}^j\;;\;\;\breve{dx}{}^i=dx^i+N^idt.
\ee

We have also used the conformal invariant variables of Lichnerowicz that
are convenient for studying the problem of initial data and the 
Hamiltonian dynamics \cite{yk,mtw}:
\be\label{lich}
 F_c^{(n)}=F^{(n)}\; || {}^{(3)}g||^{n/6}; ~~\vh_c=\mu \;|| {}^{(3)}g||^{1/6}
\ee
\be\label{civ} N_c=N\;|| {}^{(3)}g||^{-1/6},~~~~
 g^c_{ij}={}^{(3)}g_{ij}\;|| {}^{(3)}g||^{-1/3}
;~~~~~~~(||g^c||=1)
\ee
where $F^{(n)}$ is a matter field of the conformal weight $n$.

Then, the Einsten-Hilbert action assumes the form
\be\label{eha}
W^{E-H}=\int d^4x \sqrt{-g}[-\frac{\mu^2}{6}{}^{(4)}R]=\int d^4x
\left[-N_c\frac
{\vh_c^2}{6}{}^{(4)}R+\vh_c\partial_\mu(N_c\partial^\mu\vh_c)\right]
\ee
which coincides with the action of the conformal invariant
Penrose-Chernikov-Tagirov (PCT) theory of a scalar field $\Phi$
in terms of the Lichnerowicz variables (\ref{lich}), (\ref{civ})
($\vh_c= \Phi||^{(3)}g||^{1/6}$)

However, in contrast with GR the observables in PCT theory are conformaly
invariant quantities,
in particular, an observer measures the conformally invariant interval
\be\label{dscut}
(ds_c)^2=N_c^2dt^2- g_{ij}^c\breve{dx}{}^i
\breve{dx}{}^j.
\ee

The PCT version is preferable from the point of view of unification of
gravity with other interactions.
There is a possibility to identify the PCT scalar field with the modulus of
the Higgs doublet and to add the matter fields as the conformal invariant part
of the Standard Model for strong and electro-weak interactions \cite{pr,scalar}.
The obtained model was called the Conformal Unified Theory (CUT)
\cite{grg,pr,plb}.

In the first - order formalism the actions of both theories can be represented
in the form which is a continual local generalization of the extended
mechanics:

\be\label{espct}
W^{ES}[P_f, f; P_{\vh}, \vh_c]=\int\limits_{t_1}^{t_2}dt\hskip-.05cm\int
d^3x\left[ \sum\limits_{f=g_c,F} P_fD_tf-P_{\vh}D_t\vh_c -N_c
(-\frac{P_\vh^2}{4}+  {\cal H}_f)\right],
\ee
where $D_tf, D_t\vh_c$ are the time covariant derivatives
$$
D_t\vh_c=\ik_t\vh_c-\ik_k(N^k\vh_c)+{2\over 3}\vh_c\ik_kN^k.
$$

Action (\ref{espct}) is invariant under reparametrization of the
coordinate time $t$.  According to the considered gaugeless
Hamiltonian reduction, we should pick out the superfluous variable of the
extended system, which plays the role of the evolution parameter for the
corresponding reduced system.

Our idea is to carry out the global Hamiltonian reduction.
This means extracting a global component from the local scalar field as the
evolution parameter of RS
\be\label{globalphi}
\vh_c(t,x)=\vh_0(t) {a}(t,x);
~~D_t\vh_c=\frac{d\vh_0}{dt}{a}+\vh_0 D_ta.
\ee
We shall extract also a
global component from the conformal invariant lapse function
\be\label{lapse}
N_c(t,x)= N_0(t){\cal N}(t,x ).
\ee

As we have introduced one more variable, we should impose one more constraint
and take the constraint
\be\label{cgr}
\int{ d^3x} a \frac{D_t a}{N_c}=0,
\ee
which diagonalizes the kinetic part of the Lagrangian.

New variables require the corresponding momenta $P_0$ and $P_a$.
We define the decomposition of the old momentum $P_\vh$ over the new momenta
$P_0$ and $P_a$
\be\label{pphi}
P_\vh \buildrel{\rm def}\over= P_0\frac{{a}}{{\cal N}V_0}
+ \frac{1}{\vh_0}P_a ;
\;\;\;\;\;\;
V_0=\int d^3x\frac{a^2}{{\cal N}}.
\ee
to get the conventional canonical structure for the new variables
$$
\int d^3xP_\vh D_t\vh_c =P_0\frac{d\vh_0}{dt}+\int d^3x{P}_aD_t{a}.
$$

The substitution of these definitions into the old action produces a
superfluous momentum term:

\be\label{espct2}
W^{ES}=\int\limits_{t_1}^{t_2}dt\left(\int
d^3x\sum\limits_{f=a, g_c,
F}P_fD_tf-P_0\dot\varphi_0-N_0\left[-\frac{P_0^2}{4V_0}
+H_f\right]\right)
\ee

Finally, the obtained action (\ref{espct2})
acquires the structure of the extended cosmological model with two functionals:
the functional of the volume $V_0$ (\ref{pphi}), and
the functional of the Hamiltonian:
$$
H_f[\varphi_0]=
\int d^3x{\cal N}[-\frac{{P}_a^2}{4\varphi_0^2} +{\cal H}_f].
$$

Resolving the constraint

$$
\frac{\delta W^{ES}}{\delta N_0}=0\,\Rightarrow\,
(P_0)_{\pm}=\pm2\sqrt{V_0 H_f }
$$
we get the reduced action

\be\label{rspct}
W^{RS}_{(\pm)}=\int\limits_{\varphi_1=
\varphi_0(t_1)}^{\varphi_2=\varphi_0(t_2)}
d\varphi_0 \left\{\left(\int
d^3x\sum\limits_fP_fD_{\varphi}f\right)\mp2\sqrt{V_0 H_f}\right\}.
\ee

The variation of the RS action determines the dependence of all variables on
the evolution parameter
$P_f=P_f(\vh_0), ~~ f=f(\vh_0)$
and completely reproduces the Einstein equations.

Proper time dynamics gives the integrals of motion as functionals of field
variables \cite{grg},
and determines the dependence of the global part of the proper time on
the global evolution parameter:
\be\label{ptdpct1}
\frac{\delta W^{ES}}{\delta N_0}=0\,\Rightarrow\,
(P_0)_{\pm}=\pm\sqrt{V_0 H_f(\vh_0)};
\ee
$$
\frac{\delta W^{ES}}{\delta P_0}=0\,\Rightarrow\,
\left(\frac{d\varphi_0}{dT}\right)_{\pm}
=\frac{(P_0)_{\pm}}{2V_0}
=\pm\sqrt{\rho(\vh_0)},
$$
where
$$
dT=N_0dt;~~~~~~\rho =\frac{H_f}{V_0};
$$
\be\label{ptpct}
T(\vh_0)_\pm=\pm\int\limits_0^{\vh_0}d\vh\rho^{-1/2}.
\ee 
It is easy to check that the local part of the scalar field in  perturbation
theory is nothing but the potential of the Newton interaction.  To reproduce
the homogeneous FRW cosmology, it is sufficient to neglect this interactions.

Finally, we get the Hubble law in two versions, GR and CUT.

We emphasise that the definition of observables in General Relativity
(\ref{dsgr}) has some defects
as compared with CUT (\ref{dscut}).

i) In contrast with CUT, in GR there is mixing of the evolution parameter with
metric.

ii) For a space with positive curvature the Friedmann time violates causality
\cite{kpp,w} but the conformal one does not!

iii) In CUT, an observable 3D-volume is an integral of motion; in GR we have a
singularity at the beginning.

Therefore, it is worth to thoroughly consider physical consequences
of the Conformal Unified Theory

\section{Conformal World}

\subsection{Hoyle-Narlikar cosmology}

The convention of measurable interval in CUT (\ref{dscut}) leads to the
Hoyle-Narlikar cosmology \cite{N} where the
red shift is explained by the evolution
of all masses rather than by the expansion of the Universe.
A photon emitted by
an atom on a star remembers the size of this atom, and
after billions of years, the wavelength of this star photon
can be compared with that of a
 photon emitted by a standard atom on the Earth, the size of which
is much less than the size of the star atom. As result, we
arrive at the red shift

$$
Z=\frac{\vh_0(T-{D/c})}{\vh_0(T)}-1\simeq
({D/c})   {H_{\rm Hub}}(T)
+\dots
$$
with the Hubble parameter determined by the proper time dynamics
(\ref{ptdpct1})-(\ref{ptpct})
\be\label{hcut}
H_{\rm Hub}=\frac{1}{\vh_0(T)}\frac{d\vh_0(T)}{dT}=
\frac{\sqrt{\rho}}{\vh_0(T)}.
\ee

>From the last equation (\ref{hcut}) it follows that
$$
\vh_0 =
\frac{\sqrt{\rho}}{H_{\rm Hub}}.
$$
The substitution of the observational data
$$
\rho=\rho_{\rm critical}{\Omega_{\rm exp}};~~~~~~ \rho_{\rm critical}=
{3}M_{\rm Pl}^2H_{\rm Hub}^2/{8\pi}
$$
 leads to the present - day value of the scalar field which coincides with the
Newton constant in the action (\ref{esgr}) and (\ref{eha})
$$
\frac{\vh_0^2(T=T_0)}{6}=\frac{M_{\rm Pl}^2}{16\pi}\Omega_{\rm exp}\equiv
\frac{\mu^2}{6}\Omega_{\rm exp}
$$
in agreement with  astrophysical data \cite{prd}.
$$
0.02< \Omega_{\rm exp}<2.
$$

\subsection{Cosmic Higgs vacuum}

It is clear that the present - day state of the Universe
provides the laboratory vacuum.
One can only suppose that at the present day stage the total Hamiltonian
can be divided into the part
forming evolution of the Universe (the Universe expectation value) and the
Laboratory part (for which the expectation value equals zero)
$$
H_f[\varphi_0]
\buildrel{\rm
def}\over
=
\rho_{\rm Un}V_0+(H_f-\rho_{\rm Un}V_0)=
\rho_{\rm Un}(\varphi_0)V_0+H_L;
$$
$$
<Universe|H_f|Universe>
=\rho_{\rm Un}V_0;
$$
$$
<Universe|H_L|Universe>=0.
$$
It was known that
$\rho_{\rm Un}\sim 10^{79}m_{\rm proton}$ while $H_L\sim 1m_{\rm proton}$,
therefore, we can apply the nonrelativistic type approximation for decomposing
the square root over the inverse volume

$$
\sqrt{V_0H_f}=\sqrt{V_0^2\rho_{\rm un}+V_0H_{\rm lab}}=
2V_0\sqrt{\rho_{\rm un}}
+\frac{H_{\rm lab}}{\sqrt{\rho_{\rm un}}}.
$$

Finally, we get a splitting of the action on the cosmological and
laboratory parts
$$
W^R_{(+)}(p_f, f|\bar\varphi_0)=W^G_{(+)}(\bar\varphi_0)
+W^L_{(+)}(p_f, f|\bar\varphi_0).
$$

The proper time dynamics

$$
\frac{d\vh_0}{\sqrt{\rho_{\rm un}}}=dT
$$
allows us to rewrite the laboratory action in terms of the observable time
\be\label{lab}
W^L_{(+)}(p_f, f |\bar\varphi_0)=\int\limits_{T_1}^{T_2}dT\left(\int
d^3x\sum\limits_fp_fD_{T}f-H_L(p_f, f |\varphi_0(T))\right).
\ee

In this case, the laboratory part of the total action in terms  of the
conformal time coincides with the $\sigma$ model version of SM (within the
gravitational interactions).

As the time of the laboratory experiment $\Delta T$ is much smaller than the
age of the Universe $T_0$:
$$
T_1=T_0-\frac{\Delta T}{2}\,;~~\,T_2=T_0+\frac{\Delta T}{2} \,
;~~\,{\Delta T}\ll T_0,
$$

we can neglect the change of the scalar field
in laboratory experiments:
$$
\varphi_0(T)\approx\varphi_0(T_0)=M_{Pl}\sqrt{\frac{3}{8\pi}};
~~~~ T_0-\frac{\Delta T}{2} <T< T_0+\frac{\Delta T}{2}.
$$

In perturbation theory, corrections to the scale factor $a$ represent
the Newton potential
$$
\vh_c(T,x)=\vh_0(T) \;a(T,x);\;~~~~a(T,x)=1 + {\rm  (Newton\; potential)}.
$$
like in QED the time component of the electromagnetic field gives the
Coulomb potential. We see that the particle - like excitations of the scalar
field are absent, as predicted in \cite{pr}.

\subsection{$\sigma$-model version of SM}

Theoretically, the status of
the Higgs sector in conventional SM is still mysterious.
The physical motivation for its existence as
a consequence of the first symmetry principles
is unclear. The imaginary mass in the Higgs potential
is rather unexpected. There is also a number of
difficulties caused by the scalar
mode of a nonvanishing vacuum expectation value in
cosmology (a great vacuum
density \cite{1}, a monopole creation \cite{2},
domain walls \cite{3}).

Almost all we know about the Higgs
particle  comes from collider experiments.
In direct search the Higgs particle is looked
in the process
$$
e^+e^- \to Z^{(\ast)} \to Z^\ast H \to Xf\bar f.
$$

The search for the Higgs particle is a main
motivation for building  new  high energy accelerators.
The LHC will be able to give a definite answer to
the question concerning the existence of the SM Higgs
particle. Some information can be derived also from
the upgraded Tevatron. The main Higgs production
mechanisms at hadron colliders are the following:

1) $gg \to H$ \hskip1cm {\sl gluon -- gluon fusion}

2) $WW (ZZ) \to H$ \hskip1cm {\sl weak boson fusion}

3) $q\bar q \to W(Z) + H$ \hskip1cm {\sl association with W/Z}

4) $gg(q\bar q) \to t\bar t + H $ \hskip1cm {\sl association
with $t\bar t$}
\smallskip

There arises a fundamental experimental problem
how to discriminate between the SM
and an effective sigma model obtained from the Conformal
Unified Theory considered here \cite{part2}.
Some perspectives
that open with the new HE accelerators, mainly LHC,
 are in principle well known and
are widely presented in the literature \cite{LHC}
but never in the present context.

A new class of tests is
the analysis of properties of the virtual
Higgs (or rather a UV-regulator) that can be compared on different
experimental energy scales.

Our
proposition is based
on the observation that the role of the Higgs particle
as a regularizing parameter of the model can be played
by an effective parameter that is in principle dependent
on the energy of the considered process as we are working
with the $S$-matrix with a finite interval of time (\ref{lab}). This
assumed energy dependence can be tested in
high precision experiments planned in the nearest future \cite{pps2}.

\subsection{Beginning}

In the beginning of the Universe (at $\vh_0=0$) we could not separate
 the evolution of the scalar field  (or the scale factor in GR) from
 the evolution of matter fields.

To study the beginning stage, we need the classification of Hamiltonians,
which follows from the classification of times, considered in the beginning of
the present paper. This classification is given in Table 1.
\medskip
\bc
\bt{|p{7cm}||p{7cm}|}
\hline
\vskip.2cm\hskip.7cm{\bf Three Times} &
\vskip.2cm\hskip.7cm {\bf Three Hamiltonians} \vskip.1cm\\
\hskip.7cm COORDINATE TIME & \hskip.7cm   CONSTRAINT\\
\hskip.7cm\fbox{$t\to t^{\prime}=t^{\prime}(t)$} & \hskip.7cm   \fbox{$H^{ES}=0$} \vskip.1cm\\
\hskip.7cmEVOLUTION PARAMETER & \hskip.7cm   EVOLUTION HAMILTONIAN\\
\hskip.7cm \fbox{$\vh_0$}  & \hskip.7cm  \fbox{$P_0=2\sqrt{V_0H_f}$} \vskip.1cm\\
\hskip.7cm"MEASURABLE" TIME& \hskip.7cm  "MEASURABLE HAMILTONIAN"\vskip.1cm\\
\hskip.7cm{\large\fbox{$\displaystyle  dT=N_0dt$}} &  \hskip.7cm
{\large\fbox{$\displaystyle H^{M}_{(hj)}=-\frac{\partial W^{RS}}{\partial T}$}}
\vskip.2cm\\
\hline
\et\ec
\bigskip
\bc
Table 1. Classification of Hamiltonians
\ec
\bigskip

The energy constraint corresponds to the noninvariant nonobservable coordinate
time.

Superfluous momentum on the constraint can be called the evolution
Hamiltonian.

Using the Hamiltonian-Jacobi prescription we can also introduce the
"measurable Hamiltonian" as the derivative of the action with respect
to the "measurable" time.

We have considered the beginning of the Universe in the QFT approximation.
This means:

i) neglecting all interactions

ii) the decomposition over the inverse volume, and

iii) the use of oscillator-like variables (as in QFT).

\bigskip

Details of our approximation are the following.

We considered only photons and gravitons and, finally, got the set of
oscillators ($K$ stands for photons; $L$ for gravitons):

$$
 \int\limits_{V_0}d^3x
P_{(A)}^j\dot A_j^\perp=\sum\limits_{K}p_K\dot q_K;
$$
$$
\int\limits_{V_0}d^3x
P_{(h)}^{ij}\dot h_{ij}^\perp=
\sum\limits_{L}p_L\dot q_L - \frac{\dot \vh_0}{\vh_0}\sum\limits_{L}
 p_Lq_L;
$$

$$
\int d^3x\frac{1}{2}\left(P_{(A)}^2+(\ik_iA^\perp)^2\right)
=\sum\limits_{K=(k,\alpha)}\frac{1}{2}\left(
p{}_K^2+\omega_K^2q_K^2\right);
$$

$$
\int
d^3x\left(\frac{6P_{(h)}^2}{\vh_0^2}+\frac{\vh_0^2}{24}
(\ik_ih^\perp)^2\right)
=\sum\limits_{L=(l,\alpha)}\frac{1}{2}\left(
p{}_L^2+\omega_L^2q_L^2\right).
$$

Gravitons differ from photons by an additional (Hubble - like) term in the
action with a singularity at the beginning.

We get the extended system action:
$$
W^{ES}=\int dt \left[\sum\limits_{I}\dot q_Ip_I -
\dot\vh{}_0(P_0+\sum\limits_{L}\frac{q_Lp_L}{\vh_0})+
\frac{d}{dt} \left(\frac{\vh_0P_0}{2}\right)+N_0 H^{ES}\right]
$$
where
$$
H^{ES}=\left(-\frac{P_0^2}{4V_0}  +\sum\limits_{I=K,L}\frac{1}{2}(
p{}_I^2+\omega_I^2q_I^2)\right)
$$

We can in detail consider only one mode of gravitons with a wavelength
of an order of the background cosmic radiation one.

Appart from, the conventional stage of radiation of the Universe we got also
the stage of "inflation" with respect to the conformal time (see Appendix B.).

The measurable energy for photons completely coincides with the conventional
energy of photons in QFT in the flat space-time.

The measurable energy for gravitons in the present - day asymptotics coincides
with the Tolman definition of the energy momentum  tensor \cite{t}. Due to the
Hubble - like term, the measurable energy can become zero at the point where
 $\vh_{(b)}=\vh_0(T_0=0)\neq 0$ before singularity, and then accepts negative
 values.

This means that an observer  at this moment sees the creation of the Universe
>from nothing with the set of non-zero quantum numbers, integrals of motion.

Negative values can be treated correctly only in quantum theory where they can
be removed by replacing the creation to annihilation ones and vice versa.

\section{Conclusions}

We considered three new ideas related to

i) gaugeless Hamiltonian reduction of General Relativity with the internal
 evolution parameter,

ii) classification of "times", and

iii) resolving the global energy constraint.

These three ideas give us a chance to convert GR into the Conformal
Unified Theory with the set of predictions:

a) the Hoyle-Narlikar cosmology,

b) the cosmic Higgs effect for formation of masses of elementary particles
and the Newton constant of gravity,
which are defined by the
cosmological data on the density of matter and the Hubble parameter
 (the Mach principle),

c) the $\sigma$-model version of SM without the Higgs particle-like
excitations,

d) the inflation stage as consequence of the graviton dynamics at the
 beginning of the Universe.

The Lagrangian of CUT does not contain any dimensional parameters,
as masses of all fields are changed by a scalar field (multiplied by
the corresponding dimensionless coupling constants). The scalar field in CUT
restores the conformal symmetry of the action, like the vector gauge fields
restore the gauge symmetry. In accordance with the ``gauge ideology'' of Weyl,
Yang--Mills, Utiyama and Kibble, the minimal conformal - invariant dynamics
of the scalar field corresponds to the PCT action with negative sign.
This action, the scalar field, the conformal invariant variables and
measurable time can be based in the scalar version of the Weyl geometry
of similarity where we can measure only the ratio of lengths of two vectors
at the same point.

In the conventional approach, one tries to describe large-scale phenomena by
 the theory with the Higgs spontaneous symmetry breaking mechanism
invented to describe physics of elementary particles.  Here we were trying to
describe the generation of elementary particle masses by the conformal
version of the Einstein theory
proposed to describe large-scale phenomena including generation of the
Universe.

\vspace{0.5cm}

{\bf Acknowledgments}

\vspace{0.5cm}

We acknowledge interesting and critical
discussions with Profs. B.M. Barbashov, A.V. Efremov,
G.A. Gogilidze, V.G. Kadyshevsky, E. Kapu\'scik,
A.M. Khvedelidze, D. Mladenov,
V.V. Papoyan, Yu.G. Palii and G.M. Vereshkov.
One of us (V.P.) thanks Profs. C. Isham, T.W.B. Kibble, M.J. Lavelle,
M.A.H. MacCallum, D. McMullan, K. Stelle
and A.A. Tseytlin for valuable discussions and worm hospitality during his
visit to the U.K.
We also thank
the Russian Foundation  for Basic
Research, Grant N 96\--01\--01223,
the Polish
Committee for Scientific Researches, Grant N 603/P03/96
and the Bogolubov-Infeld Program for support.

\vspace{0.5cm}

\appendix

{\Large\bf Appendix A: Fermions in CUT}

\vspace{0.5cm}

\renewcommand{\theequation}{A.\arabic{equation}}

\setcounter{equation}{0}
We consider the Fock action for $(n)$ fermions \cite{fock} $\Psi^I(I=1,2,...n)$
\be
W=\int d^4x\sqrt{-g}[-\bar\Psi^Ii\gamma^\sigma
(D_{\sigma})\Psi^I-\Phi\bar\Psi^I\hat X^{IJ}\Psi^J].
\ee
where  $D_{\sigma}$ is the Fock covariant derivative, $ \hat X^{IJ}$ is the
matrix of dimensionless coupling constants in the unitary gauge of SM and
$\Phi$ is the modulus of the Higgs doublet field.

In terms of the Lichnerowicz variables $\Psi_c$, $\vh_c$ and $g^c$
(\ref{lich}), (\ref{civ}) and for the triad form of the 
Dirac-ADM parametrization
\be\label{tr}
g^c_{ij}=\omega_{i(l)}\omega_{j(l)}; 
~~~~~~\omega_{i(l)}\tilde\omega^j_{(l)}=\delta^j_i;
~~~~~(||g^c||=1)
\ee
the Fock action is
\be
W^F=\int d^3xdt[\frac{1}{i}\bar\Psi^I_c\gamma^{(0)}D_t\Psi^I_c+
J^{[kl]}(D_t\omega)_{(k)(l)}
-N_c{\cal H}_\Psi
+\partial_k(J^{(k)}N_c)]
\ee
where

\be
{\cal H}_\Psi=\vh_c\bar\Psi^I_c\hat X^{IJ}\Psi^J_c
-[i\bar\Psi^I_c\gamma_{(i)}D_{(i)}\Psi^I_c
+J^0_5 b- \partial_kJ^k]
\ee

\be\label{d}
D_t\Psi^I_c=(\partial_0-
N^k\partial_k+\frac{1}{2}\partial_lN^l)\Psi^I_c;~~~~~~
D_{(i)}\Psi^I_c=[\partial_{(i)}-\frac{1}{2}
\partial_k\tilde\omega^k_{(i)}]\Psi^I_c
\ee

\be\label{j}
J^{[kl]}=\frac{i}{2}(\bar\Psi_c^I\gamma_5
\gamma^{(j)}\Psi_c^I)\ve^{(j)(k)(l)};~~~~~~~
J^{(k)}=\frac{i}{2}
\bar\Psi_c^I\gamma^{(k)}\Psi_c^I;~~~~~~~
J^0_5=\frac{i}{2}(\bar\Psi_c^I\gamma_5\gamma^0\Psi_c^I)
\ee

\be
(D_t\omega)_{(k)(l)}=\tilde\omega^n_{(k)}[\partial_t\omega_{(l)n}-
N_iD\omega_{(l)n}+
D_{(l)}N_n];~~~~~~~~~~~
b=D_l\omega_{(i)n})
\tilde\omega^n_{(j)}\tilde\omega^l_{(k)}\ve^{(i)(j)(k)}
\ee

In the first-order formalism we have the sum of the PCT action and the Fock
one \be
W=W^{PCT}+W^F=\int dtd^3x\left[\sum\limits_{f=\vh_c,\omega,\Psi^I_c}P_fD_tf
-\frac{1}{2}\breve\partial_t(P_\vh\vh)-N_c{\cal H} + S\right]
\ee
where
\be
{\cal H}=-\frac{P_\vh^2}{4}
+\frac{3}{2}\frac{P_{\omega}^2}{\vh_c^2}]
+\frac{\vh_c^2}{6}\bar R +{\cal H}_\Psi
;~~~ S=-\frac{1}{3}\partial_j(\vh_c\partial^j(\vh_c
N_c))+\partial_j(J^jN_c);
\ee
\be
 \bar R=[^{(3)}R(g)^c)+8\vh^{-1/2}_c\partial^2
\vh_c^{1/2}];\;\;\;\;
D_t\vh_c
=\breve\partial_0\vh_c+\frac{2}{3}\partial_k(N^k\vh_c);
~~~~\breve\partial_0\vh_c=
\partial_0\vh_c-\partial_k(N^k\vh_c)
\ee
The Lagrange equations for momentum $P_f$ are
\be
P_\omega^{(k)(l)}=\frac{\vh_c^2}{6N_c}[(D_t\omega)_{(k)(l)}
+(D_t\omega)_{(l)(k)}]
+J_5^{[kl]};
\ee
\be
P_{\vh_c}=\frac{2D_t\vh_c}{N_c};~~~~~~~P_\Psi^I=\bar\Psi^I_c\gamma_0
\ee
We keep here all surface terms of the PCT actions
 and use the equality
$D_iN^i=\partial_iN^i$ for the metric with $||g^c||=1$.

The  extraction of the global components of a scalar field and the lapse
function ~(\ref{globalphi}) - (\ref{cgr}) and the Hamiltonian reduction
lead to the reduced action of the type of (\ref{rspct}) with the
time surface term:
\be
W^{RS}_{(\pm)}=
\int\limits_{\varphi_1=\varphi_0(t_1)}^{\varphi_2=\varphi_0(t_2)}
d\varphi_0 \left\{ \left(\int d^3x
\sum\limits_fP_fD_{\varphi}f\right)
\mp2\sqrt{V_0 H_f}
\pm\frac{d}{d\varphi_0}\varphi_0 \sqrt{V_0 H_f}\right\},
\ee
where $f$ runs over $a, \omega, \Psi$,
$$
H_f=\frac{H^{(-2)}}{\varphi_0^2}+ H^{(0)}+ \varphi_0 H^{(1)}+
\varphi_0^2H^{(2)}
$$

$$
H^{(-2)}=\int d^3x {\cal N}[ -\frac{P_a^2}{4}
+\frac{3}{2}\frac{P_{\omega}^2}{a^2}]
$$

$$
H^{(0)}=\int d^3x [{\cal N} {\cal H}_\Psi
+J^k\partial_k{\cal N}]
$$

$$
H^{(1)}=\int d^3x{\cal N}
a\bar\Psi^I_c\hat X^{IJ}\Psi^J_c
$$

$$
H^{(2)}=\int d^3x [{\cal N} \frac{a^2}{6}\bar R
+\frac{1}{3}\partial_j(a\partial^j(a{\cal N}))]
$$

\vspace{0.5cm}

{\Large\bf Appendix B: Inflation}

\vspace{0.5cm}

\setcounter{equation}{0}
\renewcommand{\theequation}{B.\arabic{equation}}

Consider the equation for a single graviton mode. The reduced system is
 described by two equations:
\be\label{b1}
\frac{dq_L}{d\vh_0}=\frac{p_L}{\sqrt{\rho_t}}+\frac{q_L}{\vh_0};
~~~~~~(\rho_t=\rho_0+\rho_g);
\ee
\be\label{b2}
-\frac{dp_L}{d\vh_0}=\frac{q_L\omega_L^2}{\sqrt{\rho_t}}+
\frac{p_L}{\vh_0}; ~~~~~~(\rho_g=\frac{1}{2V_0}(p_L^2+\omega_L^2q_L^2)),
\ee
where $\rho_0$ is a conserved density of the photon radiation.

The reduced system is supplemented with proper time dynamics
\be\label{ptdb}
T(\vh_0)=\int\limits_{[0]}^{\vh_0}\frac{d\vh}{\sqrt{\rho_t(\vh)}};
~~~~~~H_{\rm Hub}=\frac{1}{\vh_0}\frac{d\vh_0}{dT}=
\frac{\sqrt{\rho_t(\vh_0)}}{\vh_0}.
\ee

From comparison of two terms on the right hand side of eqs. (\ref{b1}) and
(\ref{b2}), we can see that there are two regimes:

i) $\vh_0^2\gg \rho_t/\omega_L^2$;~~~~ii) $\vh_0^2<\rho_t/\omega_L^2$.

In the first regime, we can neglect the last terms of eqs. (\ref{b1}) and
(\ref{b2}) and then we get the usual gravitational wave with conserved density
$\rho_g$ ~~($\frac {d}{d\vh}\rho_g=0$):

\be
q_L=\sqrt{2V_0\rho_g}\sin (\omega_L T+\delta_0);~~~~~~p_L=\sqrt{2V_0\rho_g}
\cos(\omega_LT+\delta_0);~~~~~~T(\vh_0)=\frac{\vh_0}{\sqrt{\rho_0+\rho_g}}.
\ee
In the second regime [neglect of the first terms
in eqs. (\ref{b1}) and (\ref{b2})] we get
\be\label{b5}
\omega q_L=\sqrt{2V_0}\frac{\vh_0}{T_0}; ~~~~~~p_L=\sqrt{2V_0}\frac{A}{\vh_0}
\ee
so that the density and the PTD are of the forms:
\be\label{b6}
\rho_t=\rho_0+\frac{A^2}{\vh_0^2}+\frac{\vh_0^2}{T_0^2};
\ee
\be\label{b7}
\vh_0(T)=\frac{T_0}{\sqrt{2}}\left[\rho_0[\cosh(\frac{2T}{T_0})-1] + 2
\frac{A}{T_0} \sinh(\frac{2T}{T_0})\right]^{1/2}.
\ee
The  density (\ref{b6})
shows that solutions (\ref{b5}) can be treated as a spontaneous
generation of space geometry due to the 3D space curvature.
From (\ref{b7}) we can see that
there is the period of inflation
in terms of the conformal time (here
measurable):
see Fig. 1. where one represents
the dependence of measurable energy
\be
Ec=2\rho_t+\sqrt{\rho_t}\frac{p_Lq_L}{\phi_0}-\sqrt{\rho_t}\frac{d}{d\phi_0}
\left(\phi_0\sqrt{\rho_t}\right)
\ee
 on the evolution parameter $\vh_0$
as the result of the numerical computations for solutions of (\ref{b1}) 
and (\ref{b2}).

\begin{picture}(40,70)
\includegraphics{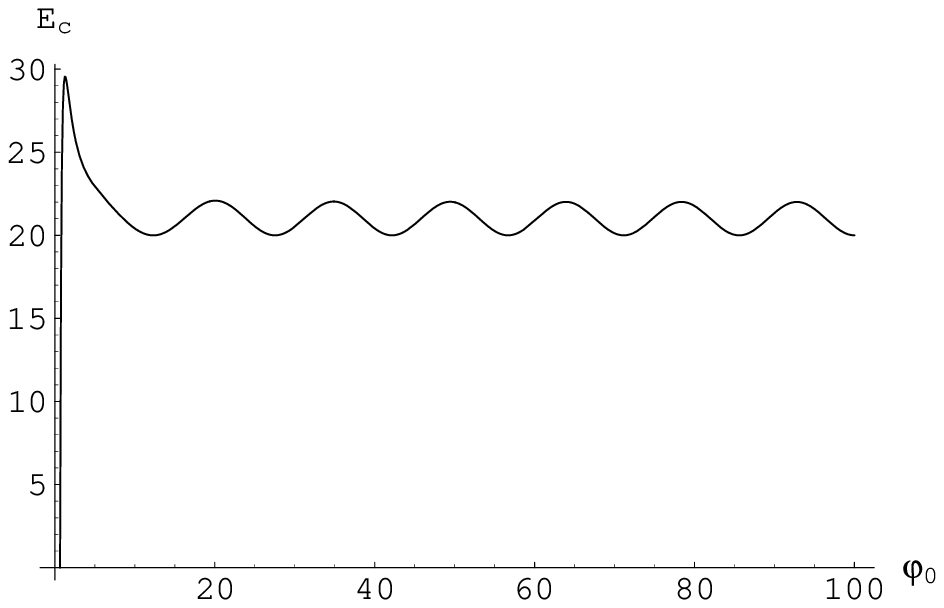}
\end{picture}

\end{document}